\documentclass[prb,showpacs,twocolumn]{revtex4}
\usepackage{graphicx,latexsym,amsmath}

\begin{document}

\title{Abrikosov vortex lattice nonlinear dynamics and its stability in
the presence of weak defects.}
\author{Alexander Yu.Galkin} \email {galkin@imp.kiev.ua}

\affiliation{Institute of Metal Physics, National Academy of
Sciences of Ukraine, Vernadskii av. 36, 03142, Kiev, Ukraine}

\author{Boris A.Ivanov}

\affiliation{Institute of Magnetism, National Academy of Sciences
of Ukraine, Vernadskii av. 36 "B", 03142, Kiev, Ukraine }

\date{\today }

\begin{abstract}
Abrikosov vortex lattice dynamics in a superconductor with weak
defects is studied taking into account gyroscopic (Hall)
properties. It is demonstrated that interaction of the moving
lattice with weak defects results in appearance of the additional
drag force, $F$, which is $F(V)\propto \sqrt V$ at small
velocities, while at high velocities $F(V)\propto 1/\sqrt V$.
Thus, the total drag force can be a nonmonotonic function of the
lattice translational velocity. It leads to the velocity
dependence of the Hall angle and a nonlinear current-voltage
characteristic of the superconductor. Due to the Hall component in
the lattice motion, the value $[dF(V)/dV]^{-1}$ does not coincide
with the differential mobility $dV/dF_{e}$, where $F_{e}$ is the
external force acting on the lattice. The estimates  suggest that
condition $dV/dF_{e} < 0$ can be met with difficulty in contrast
to $dV/dF < 0$. The instability of the lattice motion regarding
nonuniform perturbations has been found when the value of
(-$dF/dV$) is greater than a certain combination of elastic
modulii of the vortex lattice, while the more strict requirement
$dV/dF_{e} < 0$ is still not fulfilled.

\end{abstract}


\maketitle

\section{Introduction}

Vortex dynamics in superconductors is still a controversial matter
in physics of superconductivity. As a rule dynamics of Abrikosov
vortices was considered in the overdamped regime, \cite{Blater}
when the viscosity is strong and it namely determines lattice
dynamics. The Hall force was generally neglected in conventional
superconductors, where viscosity is high. Interest in the Hall
force has quickened in the past few years with fabrication of
high-$ T_c$ superclean single crystals with the large Hall
angle.\cite{Hallangle} Moreover, among other things, Hall
properties are manifested in special collective modes existing in
the vortex system. These modes were observed in resonant
experiments with $^4$He.\cite{He4} The interest to these modes as
applied to high-$T_c$ superconductors has increased considerably
in recent years in connection with experiments in BSCCO
\cite{Bismuthreson} and YBCO  \cite{Ythriumreson} compounds in
which resonant effects were observed. In YBCO system resonances
occur most likely because of cyclotronic vortex modes, while the
resonant effects in BSCCO are associated either with Josephson
plasma modes emerging there owing to Josephson coupling of
adjacent CuO$_2$ layers \cite{Bulaev} or with the vortex modes
excitations.\cite{Sonin}

The vortex modes may manifest themselves not only in resonances
but also in dissipation of the moving vortex lattice caused by
excitation of these modes. This effect is owing to irreversible
transfer of the translational vortex motion energy into elementary
bosonic excitations - the quanta of vortex modes in the presence
of the vortex-defects interaction. For the single vortex it has
been demonstrated that excitation of vortex modes results in an
anomalous behavior of the friction force, $F_v$, which is
divergent at $V\rightarrow 0$.\cite{My+PRL} The next step is study
of vortex modes excited in the moving vortex lattice.

In the present work we investigate vortex lattice dynamics in the
presence of weak defects. We are interested in an additional
effect of dissipation induced by the vortex lattice motion and
interaction with weak defects, which are not strong enough to pin
the vortex lattice, but this motion results in excitation of
normal vibrational modes and irreversible transfer of the
interaction energy to these modes (Landau damping).

Due to the additional dissipation effect nonlinearity of the
lattice motion has been revealed and it has been demonstrated that
the nonlinear regime may be even dominant one in lattice dynamics
being compared with the bare linear dissipation. When the lattice
is the subject of the Hall force we predict appearance of a
lattice motion instability. The Hall angle is found to be lattice
velocity dependent and the higher velocity the larger Hall angle
as far as it is saturated.

The paper is composed of five sections as follows. In Sec.II we
describe a microscopical model of vortex lattice dynamics with the
Hall and viscous contributions. In Sec. III we study the steady
motion of the vortex lattice and discuss the stability of this
motion. Discussion of the results of the model is presented in
Sec.IV. Finnaly, the summary is given in Sec. V.

\section{Model}

Let us consider dynamics of the vortex lattice (formed by
Abrikosov vortices parallel to the $z-$axis). Let vortices in
equilibrium to be placed in the points of an ideal lattice, $\vec
l$. Dynamics of the vortex lattice can be described on the basis
of an effective equation for the 2D vector $\vec u=\vec u_l(z,t)$,
lying in the $xy-$plane and describing the displacement of the
$\vec l$ - th vortex in the lattice (see Fig.1). \cite{Blater} The
following Lagrangian determines nondissipative dynamics of $\vec
u$:
\begin{eqnarray}
L\left\{ \vec u\right\} &=&\sum\limits_l\int dz\left[ \frac
H2\left( u_{l,x}\frac{\partial u_{l,y}}{\partial t}-u_{l,y}
\frac{\partial u_{l,x}}{\partial t}\right) \right] -
\nonumber \\
&&-W\left\{ \vec u_l\right\} -U_{imp}\{\vec u\}.  \label{lagran}
\end{eqnarray}

Here $H$ is the Hall constant of the single vortex per unit
length, $W=W\{\vec u_l\}$ is the energy of the ideal (defectless)
deformed lattice, $ U_{imp}\{\vec u_l\}$ defines the energy of
interaction of the vortex with defects (its structure will be
considered below). To describe dynamics in the linear
approximation $W\{\vec u_l\}$ should have the following structure:

\begin{eqnarray}
\label{interact} W\left\{ \vec u_l\right\}
&=&\frac{1}{2}\sum\limits_{\vec l}\int \left\{\kappa
\left[\frac{\partial \vec u_l(z,t)}{\partial
z}\right ]^2 + \right. \\
&&+ \sum\limits_{\vec l^{\prime }\neq \vec l}U^{ik}\left( \vec l
-\vec l^{\prime }\right) \delta u_{l,l^{\prime }}^i \cdot \delta
u_{l,l^{\prime }}^k \Biggr\} dz, \nonumber
\end{eqnarray}
where $\overrightarrow{\delta u}_{l,l^{\prime }}= \vec
u_l(z,t)-\vec u_{l^{\prime }}(z,t)$, $U^{ik}(\vec l )$ are
components of the force tensor which are expressed through the
second derivatives of a well-known potential of the vortex-vortex
interaction with respect to equilibrium vortex
positions,\cite{DeGennes Book}  $\kappa $ is the energy per unit
length. We neglect in the above expression of the Lagrangian the
inertial term, which comprises the vortex mass because, as was
demonstrated in \cite{My+PRL}, the vortex mass plays a substantial
role in the dissipation of the moving vortex lattice only at high
enough velocities. In the framework of a macroscopic approach,
i.e. if $\vec u_l$ varies insignificantly over a distance of the
order of the lattice constant $a_v$, the last expression
(\ref{interact}) goes over to the standard expression as with the
theory of elasticity,

\begin{eqnarray}
W\{\vec u\}&=&\int dxdydz\Biggl[\frac{c_{11}}2\left( \frac{\partial u_x}{
\partial x}+\frac{\partial u_y}{\partial y}\right) ^2+  \label{interact1} \\
&&+\frac{c_{66}}2\left( \frac{\partial u_x}{\partial y}-\frac{\partial u_y}{
\partial x}\right) ^2+\frac{c_{44}}2\left( \frac{\partial \vec u}{\partial z}
\right) ^2\Biggr],  \nonumber
\end{eqnarray}
where $c_{11},c_{44},c_{66}$ are the elastic modulii of the
lattice, $c_{44}$ can be expressed through $\kappa,\
c_{44}=\kappa/a_v^2$.

Each vortex moving in the medium experiences the action of the
drag force and the rate of its energy dissipation is proportional
to $(\partial \vec u_l $/$\partial t)^2$. For this reason, we
choose the dissipative function in the form:
\begin{equation}
\label{diss}Q=\frac \beta 2\sum\limits_l\int \left(\frac{\partial
\vec u_l}{\partial t} \right) ^2dz,
\end{equation}
where $\beta $ is the dissipation coefficient per unit length of
the vortex. When the motion of the vortex lattice is translational
and steady ($\vec u_l(z,t)=u_l(0)+\vec Vt$, where $\vec V$ is the
translational vortex lattice velocity) the dissipative function
yields the friction force per unit length of the vortex, $\vec
F=-\vec V(Q/V^2)=-\vec V\beta $.

In the case of "weak" defects the microscopical origin of defects
is not essential. \cite {My+PRL} We determine interaction between
the vortex lattice and defects in the crystal as a variation of a
local critical temperature, $T_c$, depending on coordinates. This
can be taken into account by introducing a coordinate-dependent
coefficient $[a_0+F(\vec r)]$ into the Ginzburg-Landau
expansion.\cite{Blater} We assume that the distribution of the
order parameter in the vortex does not change during the vortex
motion and is described by a known function $f(r_{\bot })$, where
$r_{\bot }^2=x^2+y^2$. \cite{DeGennes Book} Thus, regarding the
displacement of the $l$-th vortex, we can write the vortex lattice
energy $\sum\limits_l|\Psi _0|^2f(r_{l\bot }) $, where $r_{l\bot
}=[(x-l_x-u_{lx})^2+(y-l_y-u_{ly})^2]^{1/2}$ and $\Psi _{0 }$ is
the order parameter. In this case, the energy $U_{imp}$ associated
with crystal inhomogeneities can be written in the terms of the
function $f(r_{\bot }) $ in the form of a functional of the vortex
lattice displacement:

\begin{eqnarray}
U_{imp}=\sum_{\vec l}\int d\tilde xd\tilde ydzf(r_{\bot })F[\tilde
x+l_x+u_{lx}, \tilde y+l_y+u_{ly},z],  \label{functional}
\end{eqnarray}
where $\tilde x=x-l_x-u_{lx},\tilde y=y-l_y-u_{ly}$.

Finally, one arrives at the equation for the displacement of the
$l-$th vortex, $\vec u_l(z)$. It is presented in the form:
\begin{equation}
\label{equat}H\left( \tilde z\times \frac{\partial \vec
u_l}{\partial t} \right) =-\frac{\delta W\left\{ \vec u_l\right\}
}{\delta \vec u_l} -\beta \frac{\partial \vec u_l}{\partial
t}+\vec F_{l,imp}.
\end{equation}

Here the left hand side is the dynamical term governed by the Hall
(gyroscopic) constant, the terms on the right hand side are forces
acting on the $l-$ th vortex because due to interaction with other
vortices in the ideal lattice, friction and the presence of
defects.

Using the Eq. (\ref{functional}), the force caused by defects can
be written in the form of the Fourier expansion in $x$ and $y$:

\begin{equation}
\label{forcenonlin}\vec F_{l,imp}=\frac 1\Omega \sum_{\vec q}i\vec
q_{\perp }f(q_{\bot }){\cal F}(\vec q)e^{i\vec q_{\bot }\vec
l+iq_zz} e^{i\vec q_{\bot }\vec u_l(z)}\ ,
\end{equation}
where $\Omega =L_xL_yL_z$ is the superconductor volume, ${\cal
F}(\vec q)$ is the Fourier transform of the function $F(\vec r)$,
which defines inhomogeneities in the system, see
Eq.(\ref{functional}), $\vec q_{\bot }=(q_x,q_y,0),$ $f(q_{\bot
})$ is the vortex form-factor.

In the absence of dissipation and defects the equation
(\ref{equat}) can be simplified and because of translational
invariance the force tensor components
 $U^{ik}$ depend on $\vec l-\vec
l^ \prime $ only and do not depend on $z$. Then, the equation of
motion can be presented as it is done in the lattice dynamics
theory, i.e. by means of the Fourier transform of the force tensor
components. \cite {Ziman} By using the Bloch theorem it is
possible to present the equation of motion in the normal
coordinates $\vec u_{q,x(y)}$ considering that $\vec
u_{l,x(y)}=\sum\limits_{\vec q_{\bot },q_z}\vec u_{\vec
q,x(y)}\cdot \exp \left (i\vec q_{\bot }\vec l+iq_zz\right ).$

The equation of motion acquires the following form:
\begin{eqnarray}
H\dot u_{q.x} &=&C_{yy}(\vec q)\cdot u_{q,y}+C_{xy}(\vec q)\cdot
u_{q,x};
\label{eqm2} \\
-H\dot u_{q.y} &=&C_{xx}(\vec q)\cdot u_{q,x}+C_{xy}(\vec q)\cdot
u_{q,y},\nonumber
\end{eqnarray}
where $C_{ik}=\sum\limits_{l}U^{ik}(\vec l) \cdot \exp (i\vec q
\vec l \ )$ is Fourier representation of the force tensor
$U^{ik}$, see (\ref{interact}). Here and below dot  denote the
time derivative. The Fourier transform of $\vec u_l$ is discrete
in the $xy-$plane and continuous along the $z-$axis, the quantity
$\vec q_{\bot }$ has the sense of a quasimomentum, and $q_z$ is a
regular linear momentum, (we put $\hbar =1$). The quasimomentum
$\vec q_{\bot }$ there is within the first Brillouin zone, $|\vec
q|\leq 1/a_v,$ whereas $q_z$ is only restricted by the physical
reason, i.e. $q_z\leq 1/\xi ,$ where $\xi $ is the coherence
length of the superconductor. Though these values $\vec q_{\bot }$
and $q_z,$ have somewhat different physical meanings, below we
will use a macroscopical approach and retain these symbols.

The variables $\vec u_{\vec q}$ comprise the time dependent
multiplier, $ \vec u_{\vec q}\propto \exp [\pm i\omega (\vec
q)t]$, where $\omega (\vec q)$ is periodic with regard to $\vec
q_{\bot },\ \omega (\vec q + \vec g)= \omega (\vec q),\ \vec g$ is
the vector of the reciprocal lattice. Then, one can derive the
dispersion law using new parameters $ C_1=C_{xx}-\Delta
,C_2=C_{yy}+\Delta ,\ 2\Delta =\left( C_{xx}-C_{yy}\right)
+\sqrt{\left( C_{xx}-C_{yy}\right) ^2+4C_{xy}^2}$:

\begin{equation}
\label{spectrmic}\omega ^2=\frac 1{H^2}C_1C_2.
\end{equation}

At large $|\vec q |\sim 1/a_v$ the dispersion law of vortex
lattice oscillations becomes periodical in the $\vec q_{\bot }$
plane, with a period determined by the reciprocal lattice constant
and the maximal value $ q_{x,y}=1/\xi $ since the coefficients
$C_{xx(yy)}$ and $C_{xy}$ are periodical functions of $\vec
q_{\bot }$. Near the center of  the first Brillouin zone, $|\vec
q|\ll 1/a_v$, the dispersion law (\ref{spectrmic}) can be
presented via elastic moduli, $ C_1=a_v^2\left(
c_{44}q_z^2+c_{66}q_{\perp }^2\right) ,C_2=a_v^2\left(
c_{44}q_z^2+c_{11}q_{\perp }^2\right) $:

\begin{equation}
\label{spectr}\omega =\frac {a_v^2}{H}\sqrt{\left(
c_{44}q_z^2+c_{66}q_{\perp }^2\right) \left(
c_{44}q_z^2+c_{11}q_{\perp }^2\right)}\  ,
\end{equation}
which has been also obtained in Refs.\onlinecite{My+PRL, My+Fnt}.
This law corresponds to a low frequency gapless mode in the limit
of small $q_z$ and $ q_{\perp }$.

From here on we will employ a Debye-like model, with approximation
of the dispersion law by the long wave equation (\ref{spectr}) and
with taking into account periodicity in the $q_x, q_y$ directions.
The dispersion law for the vortex lattice is schematically drawn
in Fig.2. The Debye frequency for $ \omega (q_{\bot }),\ \omega
_D$, is marked in Fig.2 by a doted line.

We have concerned lattice oscillations without dissipation and
defects and found the dispersion law of normal modes. The
dissipation consideration does not make the problem more
complicated as equations are still linear. However, the presence
of defects results in nonlinear equations as the interaction
energy (\ref{functional}) and force (\ref{forcenonlin}) depend
nonlinearly on $\vec u$. A general solution of this nonlinear
equation (\ref{equat}) cannot be found. In line with \cite{My+PRL}
we can carry out a complete analysis assuming that deviations from
the uniform translational and steady motion of vortices due to
defects are small. Putting $\vec u=\vec e_xVt+\vec {\tilde u}(\vec
r,t)$ and linearizing (\ref{forcenonlin}) in respect to $\vec
{\tilde u}$, we obtain the expression for $F_{imp}$ without $\vec
{\tilde u}_l(z)$:

\begin{equation}
\label{forcelin}\vec F_{l,\ imp}=\frac 1\Omega \sum_{\vec q}i\vec
q_{\perp }f(q_{\bot }){\cal F}(\vec q)e^{i\vec q_{\bot }\vec
l+iq_zz-iq_xVt}\ .
\end{equation}

Thus, the lattice motion through the nonuniform medium results in
excitation of small oscillations of vortices in the lattice
(vortex lattice normal modes), which are described by $\vec
{\tilde u}$. It was discussed by the authors of Ref.
\onlinecite{KoshVin} in the terms of "the lattice melting".
Nevertheless, it is essential for us that oscillations bring into
appearence of the additional contribution to dissipation of the
lattice, $\Delta Q=(\beta /2)\sum_{l} \int (\partial \vec{\tilde
u}_l/\partial t)^2dz$, and in turn into the additional drag force,
$\vec F(V)$,

\begin{equation}
\vec F(V)=- \vec V\frac{F(V)}{V} ,\  F(V) = \frac{\Delta Q(V)}{V}
\ .
\end{equation}

The general equation for $\vec {\tilde u}_q$ (compare with
(\ref{eqm2})) reads
\begin{eqnarray}
\tilde u_{q,\ x} =\frac 1D\left[ \left( C_{yy} +i\beta q_xV
\right)
f_x-\left(C_{xy}-iHq_xV \right) f_y\right] ;  \label{coordinate} \\
\ \tilde u_{q,\ y} =\frac 1D\left[ \left( C_{yy} +i\beta q_xV
\right) f_y-\left(C_{xy}+iHq_xV \right) f_x\right] , \nonumber
\end{eqnarray}
where $C_{ik}\equiv C_{ik}(\vec q),\ D=(C_1+i\beta q_xV )\cdot (
C_2+i\beta q_xV ) -(Hq_xV )^2,\ f_{x,y}$ is related to
(\ref{forcelin})
\begin{equation}
f_{x,y}=\frac i\Omega  q_{x,y}f(q_{\bot }){\cal F}(\vec
q)e^{-iq_xVt}.
\end{equation}

It is clearly seen that the condition  $D = 0$ at $\beta =0 $
defines the dispersion law found in (\ref{spectrmic}), if one
would replace $\omega \rightarrow q_xV$. It is a typical evidence
of resonant character of excitation of normal modes  because of
the lattice uniform motion.

To analyze specifically a contribution of these small oscillations
to dissipation of the moving vortex lattice we assume that
inhomogeneity is caused by a system of point defects whose size is
smaller than the radius of the vortex core. In this case, the
function $ F(x,y,z)$ in (\ref{functional}) can be written as the
sum of Dirac delta functions:
\begin{equation}
\label{delta}F(\vec r)=\sum_a\alpha \delta (\vec r-\vec r_a).
\end{equation}

Here $\vec r_a$ is the coordinate of the $a-$ th defect and
$\alpha $ characterizes the intensity of interaction of the
defect. The force  acting on the $l-$th vortex with regard to the
model (\ref{delta}) is written as (\ref{forcelin}) with
substitution
\begin{equation}
\label{forcedelta} {\cal F}(\vec q)= \alpha  \sum_a e^{i\vec q\vec
r_a}.
\end{equation}

By substituting (\ref{coordinate}) with the taking into account
(\ref{forcedelta}) in the dissipative function, employing the
transformation
\begin{equation}
\label{sum-int} \frac{1}{\Omega}\sum\limits_l\int dz
\left(\frac{\partial \vec{ \tilde u}_l}{\partial t} \right)
^2\rightarrow \frac{\Omega}{(2\pi)^3a^2_v}\int d^3q(q_xV)^2\vec{
\tilde u}_q \vec{ \tilde u}_q^{*},
\end{equation}
where star denotes the complex conjugation, and carrying out
averaging over defects with the help of the relation
$\sum\limits_a\exp (i \vec q\vec r_a)=N_{imp}\delta _{q,0}$, where
$N_{imp}$ is the number of defects, $\delta _{q,0}$ is the
Kronecker symbol, we can find the additional dissipation of the
moving lattice. It may be associated with the drag force $F(V)$.
In the framework of the model (\ref{delta}) one can obtain the
equation for the additional drag force $F(V)$ per unit length of
the vortex

\begin{widetext}
\begin{equation} \label{dissipation}
F(V) =C_{imp}\frac{\beta V\alpha ^2}{(2\pi )^2}\int d^3q\cdot
q_x^2f^2(q_{\bot })
  \frac{q_{\bot }^2 \left [q_xV^2(\beta ^2+H^2)+C_2^2\right ]+
(C_1+C_2)(q_x \sqrt \Delta-q_y\sqrt
{C_1-C_2-\Delta})^2}{[C_1C_2-(\beta ^2+H^2)(q_xV)^2]^2+\beta
^2(q_xV)^2(C_1+C_2)^2}\ ,
\end{equation}
\end{widetext}
where $C_{imp}$= $N_{imp}/\Omega $ is the defect concentration.

Because of the factor $ \beta V$ appearing in (\ref{dissipation})
in front of the integral one could assume a linear velocity
dependence of the drag force, $F(V)\propto \tilde \beta V$, where
$ \tilde \beta $ is some constant. Then, it seems that a
consideration of vibrations of vortices associated with
interaction with defects results only in a small correction to the
\textit{bare} friction force, $\vec F_{bare}=-\beta \vec V$.
However, the integral in (\ref{dissipation}), as well as for the
case of the single vortex, \cite{My+PRL} contains singularities,
therefore, the additional relaxation channel can become
significant and even predominant. The most vivid example of this
fact is that the additional dissipation can be finite in the limit
of zero bare dissipation, \cite{My+PRL} $F(V)$ is finite at $\beta
\rightarrow 0$, as $F_{bare}\rightarrow 0$.

At first glance, this appears as paradoxical. However, such a
colissionless damping (Landau damping) stemming from the energy
transfer from one mode to another appears in many branches of
physics. For instance, flux-flow dynamics caused by excitations of
low frequency fermionic modes has been proposed in Ref.
\onlinecite{KopninVolovik} and the finite friction force
independent of the bare friction is predicted there. The friction
force finite at $V\rightarrow 0$ and caused by excitation of
bending oscillation of the domain wall in ferromagnets with
microscopic defects was predicted in Ref. \onlinecite{dw}.  As it
will be demonstrated below here $F(V)\propto V^{\delta},\ \delta <
1$ at $V \rightarrow 0$ and $F(V) > F_{bare}$ at any $\beta$ and
small enough velocities.

This behavior can be explained as following. The expression
(\ref{dissipation}) contains $\beta $ in the combination $ \beta
/\left\{ \beta ^2+G^2\left[ \vec q,(q_xV)^2\right] \right\} $,
where the function $G$ is such that the condition $G\left[ \vec
q,(q_xV)^2\right] =0$ with the substitution $q_xV\rightarrow
\omega $ defines the frequencies of normal modes of vortex
vibrations in the lattice (\ref{spectr}). For $ \beta \rightarrow
0$ the expression (\ref{dissipation}) is transformed into the
$\delta $-function, and becomes $\pi \delta \left\{ G\left[ \vec
q,(q_xV)^2\right] \right\} $. After simple transformations this
expression can be reduced to the $\delta $-function of the type
$\delta \left[ (q_xV)^2-\omega ^2(\vec q)\right] $, where $\omega
(\vec q)$ is the frequency of normal mode (\ref {spectr}). Since
the equation $q_xV=\omega (\vec q)$ possess a solution at any
finite velocity the value of $Q(V)$ can be finite even at $\beta
\rightarrow 0$.

In the case $\beta \rightarrow 0$, it is obvious that the
additional dissipation becomes the main source of dissipation in
the system. This may be realised if the friction force has a
dependence like $F\propto V^\delta ,\delta <1$, as it was revealed
in the case of the single vortex.

There is an alternative explanation of the additional dissipation
appearance. The modes excitation in the limit of small dissipation
can be described on the basis of momentum and energy conservation.
Let us go over to a reference frame moving with vortices. In this
reference frame defects move at a velocity $ -\vec V=V\vec e_x$
parallel to the $x-$axis and can transfer the momentum $ \vec q$
to the vortex lattice only simultaneously with the energy $\vec
q\vec V=q_xV$. This momentum is redistributed between the lattice
as a whole and an elementary excitation, and the energy is
transferred to elementary excitations only. In particular, in the
case of the single vortex the vortex as a whole acquires momentum
perpendicular to its axis ($z-$axis), while the wave propagating
along the vortex obtains the $z-$axis momentum projection.

Let us analyse the general expression (\ref{dissipation}). Note
that integration has to be done not only inside the first
Brillouin zone. The integration is limited only by cut off $\tilde
q_{\bot }$ over $1/\xi $, when the vortex form factor
$f(q_{\bot})\neq 0$. In the case of the vortex lattice, the
lattice constant $a_v$ is usually assumed to be larger than $\xi
$. In order to resolve the vortex lattice problem one should go
beyond the longwave limit. Generally speaking, one has to
integrate (\ref{dissipation}) taking into account complicated
periodical dependencies of $C_{1},\ C_{2},\ \Delta$ of the wave
vectors, and a non-periodic dependence of $f(q_{\bot })$, terms
with $q_xV$ and so on. Nevertheless, the situation is simplified
in limiting cases, for example, at small or high velocities.

First we concern the case of small dissipation, $\beta \ll H$, for
which the conservation law $\omega (\vec q)=q_xV$ defines an area
that contributes to the integral in (\ref{dissipation}). To
provide an estimation, we assume that $c_{11}\sim c_{66}\sim
c_{\bot },$ then $\omega H=a^2_v(c_{44}q_z^2+c_{\bot }q_{\bot
}^2)$, and the condition $\omega (\vec q)=q_xV$ may be written as

\begin{equation}
\label{integration}c_{44}c_{\bot }q_z^2+(c_{\bot
}q_x-VH/2a_v)^2+c_{\bot }^2q_y^2=(VH/2a_v)^2.
\end{equation}

This expression indicates that at small $V$ the size of the region
in the $ \vec q_{\bot }$ plane, which contributes to the integral
in (\ref{dissipation}) is proportional to $V$.  If one considers
that $c_{11}\neq c_{66}$, then of course the above consideration
remains true and only the expression (\ref {integration}) becomes
more complicated. Thus, at $V\rightarrow 0$ the region near $\vec
q=0$ gives rise to a very small contribution to $Q$, which is
proportional to $V^\delta,\ \delta
>1$.\cite{My+Fnt}

However, if periodicity is taken into account, then the areas with
$\omega (\vec q)\rightarrow 0$ emerge near all points in the $\vec
q_{\bot }$ plane, which are equivalent to the origin. If one puts
$\vec q_{\bot }=\vec g+\tilde q_{\bot }$, where $\tilde q_{\bot }$
is the wave vector within the first Brillouin zone and $\vec g$ is
the wave vector of the reciprocal lattice, then these points are
located at $\tilde q_{\bot }\rightarrow 0$. It is worth to mention
that values of $q_xV$ near these points are $ (g_x+\tilde
q_x)V\approx 2\pi nV/a_v$. The corresponding area in the $\vec
q_{\bot }$ plane is now given by $\omega (\vec q)=2\pi nV/a_v$.

The value of $g_x$ is limited by the cutoff $g_x\leq 1/\xi $ or
$n\leq (a_v/2\pi \xi )$. Hence, the condition of the area
smallness in the $\vec {\tilde q}_{\bot }$ plane with substitution
of $q_x \sim g_x \sim 1/a_v$ in $q_xV$ and  with $\vec {\tilde
q}_{\bot }$ in $C_1, C_2$ without changes takes the form:

\begin{equation}
\label{integration1}
 a_v^2c_{\bot }(c_{44}q_z^2+c_{\bot} \vec {\tilde q}_{\bot
 }^2)=(VH/\xi)^2.
\end{equation}

Thus, these regions can be small being compared to the size of the
first Brillouin zone at velocity $V\ll V_c$, where $V_c=\xi \bar c
/H \sim \omega _D\xi $, where $\bar c$ is some combination of
$c_{11}, c_{66}$ (more accurate expression for $V_c $ will be
given below, see Eq. (\ref{Vcrit})). A rough estimate of the
characteristic velocity $V_c$ can be done through the Debye
frequency, $V_c \sim \omega_D\cdot \xi$. The Debye frequency
appears because of periodicity of the dispersion law and the
coherence length indicates that we take into consideration all
$\vec  q_{\bot } $ wave vectors from $1/a_v$ to $1/\xi $. (Note
$V_c$ is much smaller that the characteristic phase velocity of
collective modes, $V_{ph} \sim \omega_D\cdot a_v$.)

The integral in (\ref{dissipation}) breaks into the sum over
different $\vec g$. In the framework of a macroscopical approach
the above derived formula (\ref{dissipation}) can be presented
with coefficients $C_1, C_2$ and $\Delta$ that acquire macroscopic
form: $C_1/a_v^2=c_{11}q_{\bot }^2+c_{44}q_z^2,\
C_2/a_v^2=c_{66}q_{\bot }^2+c_{44}q_z^2$ as were defined above and
$\Delta = (c_{11}-c_{66})q_y^2a_v^2$, $C_1-C_2-\Delta =
(c_{11}-c_{66})q_x^2a_v^2$. Then, the last bracket in the
numerator, after replacement $\vec q \rightarrow \vec {\tilde q}$,
is equal to zero, that simplifies integral calculation. All $\vec
q$ vectors those are not included in $C_1, C_2$ or $\Delta$ should
be replaced by corresponding $\vec g$ vectors then one can
substitute in (\ref{dissipation}) both $g_x=\left( 2\pi
/a_v\right) n$ instead of $q_x$ and $g_{\bot }=\left( 2\pi
/a_v\right) (n^2+m^2)^{1/2}$ instead of $q_{\bot }$.

The sum over $m$ and $n$ may be replaced by the integral as we
consider large $m$ and $n$, $m,n\leq a_v/\xi $ and we integrate
over $\vec {\tilde q}$ only a small region near each vector of
reciprocal lattice, $|\vec {\tilde q}|\leq V/V_c$, and considering
large enough values of $\vec g$ till $1/\xi $.  The expression for
the drag force per single vortex in the limit of small viscosity,
$H\gg \beta $, and small velocities, $V\ll V_c$, finally is

\begin{eqnarray}
&&\ F=\gamma _HV^{1/2}, \label{limit1}
\\ &\gamma _H&=\frac{ 2 \pi
\alpha ^2C_{imp}}{3
\sqrt{c_{11}c_{44}c_{66}}}\frac{J\sqrt{H}}{a_v\xi ^{3/2}}\cdot
  \begin{cases}
    3/\sqrt 2 ,&  c_{11}\sim c_{66}\\
    \left( c_{11}/c_{66}\right )^{1/4}, & c_{11}\gg c_{66},
  \end{cases}
  \nonumber
\end{eqnarray}
where $J=\int \xi ^{3/2}dg_xdg_yf^2(g_{\bot })g_x^{3/2}g_{\bot
}^2$ is the constant of the order of unity.

In the limit of high viscosity, $\beta \gg H$, and $V\rightarrow
0$ the similar estimation demonstrates that again only small areas
near $\vec g$ are important and calculations yield

\begin{equation}
\label{limit2}F=\gamma _\beta V^{1/2},\gamma _\beta =\frac{\sqrt 2
\pi \alpha ^2C_{imp}}{\sqrt{c_{44}}}\frac{J\sqrt{\beta }}{a_v\xi
^{3/2}}\left( \frac 1{c_{66}}+\frac 1{c_{11}}\right),
\end{equation}
where $J$ is the same constant as in (\ref{limit1}).

So for small velocities, $V\ll V_c$, and any relation between $H$
and $\beta $ the drag force is $F(V)=\gamma (H,\beta )\sqrt{V }$,
where the coefficient $\gamma (H,\beta )$ in limiting cases is
given by (\ref{limit1}, \ref{limit2}).

When $V\sim V_c$ one has to take into account not only small
regions of the wave vectors near $\vec g$, but perform integration
over all $|\vec q|\leq 1/\xi $ as well. But this is an extremely
complicated problem and we are not able to solve it even
numerically as the detailed behaviour of $C_{ik}$ at large $\vec
q$ is not known. Far above the critical velocity, $V\gg V_c$, one
can significantly simplify the problem. In this case, it is
possible to neglect contribution of the terms like $(c_{11},\
c_{66})q_{\bot }^2$ having the order of the value $ \omega _D$ to
the denominator in (\ref{dissipation}), whereas the terms $
c_{44}q_z^2,\ \beta q_xV,\ Hq_xV$ should be taken into
consideration. Then, the calculations give at $V>V_c$
\begin{equation}
\label{visccoef}F=\frac{\eta (H,\beta )}{\sqrt{V}},\ \eta
=\frac{\pi \alpha ^2C_{imp}J^{\prime }A(H,\beta
)}{\sqrt{c_{44}}a_v\xi ^{1/2}},
\end{equation}
where $A(H,\beta )=1\sqrt{2H}$ at $H\gg \beta $ and $A(H,\beta
)=1/(2\sqrt{ \beta })$ at $H\ll \beta $, the constant $J^{\prime
}=\int \xi ^{1/2}dq_xdq_yf^2(q_{\bot })q_x^{1/2}q_{\bot }^2\sim
J$. Below the difference between $J$ and $J^{\prime}$ will be
disregarded. The dependence $F\propto 1/\sqrt{V}$ corresponds to
the single vortex limit,\cite{My+PRL} if one would replace
$c_{44}a_v^2\rightarrow \kappa $. It is not surprising because in
the limit $V\gg V_c$ the only contribution the $q_z$ wave vector
to $\omega (\vec q)$ must be taken into account. This is exactly
the case of the single vortex, for which $q_x$ and $q_y$
components of the momentum $\vec q$ are absorbed by the vortex,
while elementary excitations propagating along the vortex axis
acquire the momentum $q_z$ and the energy $q_xV$, that does not
restrict the integration area over $\vec q$ (in contrast to
(\ref{integration}) or (\ref{integration1})).

Thus, the velocity dependence of the additional drag force, which
results from the vortex-defects interaction can be described as
$F=\gamma (H,\beta ) \sqrt{V},\ V<V_c$, and $F=\eta (H,\beta
)/\sqrt{V},\ V>V_c$, where the coefficients $\gamma $  and $\eta $
are derived in Eqs. (\ref{limit1}-\ref{visccoef}). To describe the
total friction force the bare contribution $\vec F_{bare}=-\beta
\vec V$ should be concerned. The total friction force acting on
the moving vortex lattice having the different velocity dependence
for small and large velocities can be presented as $\vec F=-(\vec
V/V)F(V)$, with the interpolating equation for $F(V)$
\begin{equation}
\label{sumdrag}F(V)=\frac{\gamma (H,\beta )\sqrt{V}}{1+V/V_c}
+\beta V,
\end{equation}
where we use $V_c=\eta (H,\beta )/\gamma (H,\beta )$ as a quantity
characteristic of the problem. The schematic picture of the $F(V)$
is presented in Fig.3.

This value does not contradict to estimates provided above in
which only the order of $V_c$ has been considered. From
(\ref{limit1}-\ref{visccoef}) one can extract the characteristic
velocity $V_c$ in limiting cases

\begin{equation}
\label{Vcrit}V_c = \frac{\xi}{2\sqrt 2} \cdot
     \begin{cases}
   (\sqrt 2/H)\sqrt{c_{11}c_{66}} , & H\gg \beta,\  c_{11}\sim c_{66},  \\
       (3/ H)\left( c_{11} c_{66}^3\right)^{1/4},  &H\gg \beta,
         c_{11}\gg c_{66},   \\
c_{11}c_{66}/[\beta \left(c_{11}+c_{66}\right)], & \beta \gg H .
      \end{cases}
\end{equation}

In the problem one more characteristic velocity emerges,
$V_{trans}$, at which the bare dissipation starts to be dominant
compared to the additional one, i.e. $ \beta V \geq \gamma _H
\sqrt{V}/\left ( 1+V/V_{c} \right )$ at $V > V_{trans}$. We would
like to emphasize that $V_{trans}$ is determined as a velocity
below that the additional drag force prevails over the bare drag
force, whereas $V_c$ is defined only by the additional drag force
behaviour. In virtue of this, $V_c$ and $V_{trans}$ depend on the
problem parameters in different manners. As it is seen from
(\ref{Vcrit}), $V_c$ does not depend on the lattice-defects
interaction intensity, i.e. does not comprise $C_{imp}$ or
$\alpha$. On the contrary, $V_{trans}$ is directly correlated with
the intensity $\alpha$. Because of the nonmonotonic velocity
dependence of the additional drag force, $V_{trans}$ exists for
any as small as one likes value of $C_{imp} \alpha ^2$ and only
the character of the total drag force, $F(V)$, is changed.

If the lattice-defects interaction intensity is small, while the
bare constant $\beta $ is large, then $V_{trans}$ falls on the
increasing part of $F(V)$ and $V_{trans}=\left (\gamma/\beta
\right)^2$. It corresponds to the case $V_{trans}\ll V_c$ and the
velocity $V_c$ does not manifest itself at all: $F(V)\propto \sqrt
V$ at $V \ll V_{trans}$ and $F(V)\propto \ V$ at $V > V_{trans}$.

In the opposite case $V_{trans} \gg V_c$, $F(V)\propto \sqrt V$ at
$V < V_c$ and $F(V) \propto 1/\sqrt V$ at $V_c \ll V \ll
V_{trans}$, while the linear asymptotic $F(V) \propto V$ appears
only at $V > V_{trans} \gg V_c$ and $V_{trans}=\left( \eta /\beta
\right) ^{2/3}$. Obviously that namely in the latter case the
nonmonotonic $F(V)$ dependence may be realised.

The concrete values of this velocity and the relation between
$V_c$ and $V_{trans}$  will be estimated below.

\section{forced motion of the lattice}

In the previous section we have employed a model of a rather weak
random force induced by defects acting on the lattice. Now we
proceed to description of the vortex lattice dynamics within a
somewhat broader macroscopical approach. Having use the
macroscopical form of the average friction force (\ref{sumdrag})
instead of the random vortex-defects force considered in
(\ref{equat}) we are able to reformulate the vortex lattice
dynamics problem on the macroscopical basis.

Consider the displacement vector of vortices in the lattice $\vec
U = \vec U(\vec r, t)$ assuming that $\vec U$ changes
insignificantly over the scale of the lattice constant. We use the
trivial averaging of the Lagrangian (\ref{lagran}) with the
dynamical part presented as in (\ref{equat}) and with
substitutions $\sum _l(...) \rightarrow \int dxdy/a_v^2(...)$ and
$\vec u_l(z,t) \rightarrow \vec U(\vec r,t)$. The energy of
lattice deformation is described by the elasticity theory
(\ref{interact1}). We assume the presence of the external force,
$F_e$, while the effective drag force is determined by
Eq.(\ref{sumdrag}) with the change $\vec V\rightarrow \dot{\vec
U}$.

The equation of motion takes the similar form as (\ref{equat}):

\begin{equation}
\label{equaVec}H\left( \tilde z\times \dot{\vec U}\right)
+\dot{\vec U}\frac{F(|\dot{\vec U}|)}{| \dot{\vec U}|}
+\frac{\partial W\{ \vec U\} }{\partial \vec U}=\vec F_e.
\end{equation}

Using the equation (\ref {equaVec})as the base we consider
macroscopical dynamics of the vortex lattice, in particular, its
stability against small perturbations. We present translational
lattice velocity, $\dot {\vec U} = V$, as a sum of the steady
velocity of the lattice, $\vec V_0$, and the small correction
$\dot {\vec u},\ \dot{\vec U}$ $=\vec V_0+\dot{\vec u}(\vec r,t)$.

First we seek the solution for the lattice steady velocity $ \vec
V_0$. It allows us revealing $F_e(V)$ dependence, which is merely
current-voltage ($I-V$) characteristic of the superconductor, see
Ref.\onlinecite{Tinkham}. Because of the Hall term in (\ref
{equaVec}) the steady velocity is convenient to express via
components $\vec V_0=V_{\parallel }\vec e_{\parallel }+V_{\perp
}\vec e_{\bot },$ where $\vec e_{\parallel }=\vec F_e/|\vec F_e|,\
\vec e_{\bot }= (\hat{\vec z} \times \vec e_{\parallel })$. To
describe the stationary flow of the vortex lattice with velocity $
\vec V_0$ a couple of equations of motion are found,

\begin{equation}
HV_{\perp }+\frac{F(V_0)V_{\parallel }}{V_0} =F_e,\
 -HV_{\parallel }+\frac{F(V_0)V_{\perp }}{V_0}=0.
\label{steadyMotion}
\end{equation}

From these equations one can derive the components of stationary
velocity $ V_{\perp }$ and $V_{\parallel }$ via $V_0$:\

\begin{equation}
\label{components}V_{\perp }=\frac{F_e/H}{1+\left( F/HV_0\right)
^2},\ V_{\parallel }=\frac{FF_e/H^2V_0}{1+\left( F/HV_0\right)
^2}.
\end{equation}

For simplicity we use $F^{\prime }=[dF(V)/dV]\bigl|_{V=V_0 }
\bigr. ,\ F=F(V_0)$. With the help of (\ref{components}) the Hall
angle could be given as

\begin{equation}
\label{Hallangle}\tan \alpha =\frac{V_{\perp }}{V_{\parallel}}
=\frac{HV_0}{F(V_0)}.\
\end{equation}

So if the drag force is a linear function of the translational
velocity $V_0, \  F = \beta V_0$ (the usual case), then $\tan
\alpha $ does not depend on the velocity and the Hall angle
tangent is merely the ratio of the Hall constant and the
dissipation constant $\beta $. However, by considering the
specific drag coefficient in Eq.(\ref {sumdrag}) a striking
feature appears, namely a velocity dependence of the Hall angle.
In the limit $V\rightarrow 0\ \tan \alpha =\sqrt{V_0}/\gamma
(\beta, H)$, and the higher the lattice velocity the Hall angle
larger, then it is saturated above the characteristic velocity
$V_{trans}$.

By solving the Eq.(\ref{components}) with the taking into account
the drag force from Eq.(\ref{sumdrag}) the dependence of $F_e(V)$
acquires the following form:
\begin{equation}
\label{extervel} V_0^2H^2+\left[ \frac{\gamma \sqrt{V_0}}{
1+V_0/V_c}+\beta V_0\right]^2 \ =F_e^2
\end{equation}

The Eq.(\ref{extervel}) has a simple form of the vector sum of two
forces: the Hall force (the first term on the left hand side) and
the drag force (the second term). The differential mobility
related to the external force is
\begin{equation}
\label{diffmobil} \frac{dV_0}{dF_e}=
\frac{F_e}{V_0H^2+F(V_0)F^{\prime}(V_0)}.
\end{equation}
If one puts $H = 0$ in (\ref{diffmobil}) then the differential
mobility $dV_0/dF_e$ coincides with  $(dF/dV_0)^{-1}$. But for
$H\neq 0$, especially at $H \gg F(V)/V$ these two quantities
differ substantially.

The negative value of the differential mobility is usually
considered as a condition (both necessary and sufficient) of the
instability appearance. Nevertheless, it is quite possible that
$dV_0/dF < 0$, whereas $dV_0/dF_e > 0$, see (\ref{diffmobil}).
Therefore, one has to reconsider the instability criterion for the
case of macroscopic  dynamics in the presence of gyroforce. This
problem is of interest not only  Abrikosov lattice dynamics, but
it is also important for dynamics of any system having gyroforce
and a nonmonotonic velocity dependence of the drag force, $F(V)$
(for instance, vortices in He$^4$, Bloch lines, etc.)

To investigate the stability of the vortex lattice we take $ \vec
U =\vec V_0t+\vec u(\vec r,t)$ and find the solution in the form
$\vec u(\vec r,t)\propto \vec u_q \exp (\Lambda t+i\vec q \vec r
)$, where $\Lambda $ defines the character of time evolution of
small deviations from the steady motion. If the real part of
$\Lambda$ is negative, the value of $-\text {Re}\Lambda > 0$
serves as a damping coefficient. Otherwise, if the real part of
$\Lambda$ is positive for some values of $\vec q$, corresponding
small deviations cause the motion instability with the instability
increment Re$\Lambda>0$.

For the small correction $\vec u$ the couple of equations of
motion are

\begin{eqnarray}
&u_{q,x}\left[ \Lambda F^{\prime }(V)+C_{xx} \right]
+u_{q,y}\left(
C_{xy}-\Lambda H\right) =0;  \label{corrMotion} \\
&u_{q,x}\left( C_{xy}+\Lambda H\right) + u_{q,y}\left[ \Lambda
F(V)/V+C_{yy}\right] =0,  \nonumber
\end{eqnarray}
where $C_{ik}\equiv C_{ik}(\vec q)$ are the coefficients used in
(\ref{coordinate}), strictly speaking, their presentation in the
longwave limit, $V$ is the translational velocity of the lattice
(for simplicity we omit here and futher the index $0$). It is more
convenient to use in Eq.(\ref{corrMotion})  $\vec e_x,\ \vec e_y$
instead of $\vec e_{\|},\ \vec e_\bot$. It is necessary to stress
that this equation contains $F(V)$ and $F^{\prime}(V)= dF(V)/dV$,
where $F$ is the total drag force, rather than differential
mobility $dF_e/dV$. As it will be demonstrated below, this is a
very essential point which leads to the fact that the instability
condition is not related directly to the usual one, $dF_e/dV < 0$.

The characteristic equation for $\Lambda$ takes the form:
\begin{eqnarray}
&&\Lambda ^2\left[ F^{\prime } (F/V)+H^2\right] +\omega ^2(\vec
q)H^2+
\label{LambdaEq} \\
&&+\Lambda \left[  F^{\prime } C_{yy}+(F/V)C_{xx}\right] =0,
\nonumber
\end{eqnarray}
where we used the equation $C_{xx}C_{yy}-C_{xy}^2=\omega ^2(\vec
q)H^2$, $\omega (\vec q)$ is the dispersion law for free vortex
oscillations (\ref{spectrmic}) obtained above. The equation
(\ref{LambdaEq}) possess two solutions:
\begin{equation}
\Lambda = - \Gamma (q)\pm i\Omega (q), \label{LambdaM}
\end{equation}
where
\begin{eqnarray}
\Gamma(q)&=&\frac{(F/V)C_{xx}+C_{yy}F^\prime}{H^2+(F/V)F^{\prime }
}, \label {LambdaRoot}\\
 \Omega^2(q)&=&\frac{H^2\omega^2(q)}{H^2+(F/V)F^\prime}-\Gamma^2(q).
 \nonumber
\end{eqnarray}

In order to get the instability condition we analyse
Eqs.(\ref{LambdaM}, \ref{LambdaRoot}). If $F^{\prime }>0$, then
$\Gamma >0$ and small oscillations decay for all values of $\vec q
$. These are small damped oscillations with the frequency $\omega
\approx \Omega (\vec q)$ at $\Omega \ll \Gamma$, otherwise, for
$\Omega \gg \Gamma$, this is the case of overdamped dynamics of
$\vec u$. Thus, the inequality $F^{\prime} <0$ is the
\emph{necessary} condition for the lattice instability.

But if $F^{\prime} <0$, it is not \emph{sufficient} to induce the
lattice instability. The further analysis of Eqs.(\ref{LambdaM},
\ref{LambdaRoot}) reveals two different and independent sufficient
conditions imposed by the instability requirement $\text
{Re}\Lambda
>0 $.

The first condition is $H^2+(F/V)F^\prime <0$ or

\begin{equation}
\label{NC1} -F^\prime  (F/V)>H^2.
\end{equation}

This requirement is equivalent to the inequality $dF_e/dV < 0$,
i.e. the rise of the negative differential mobility, in a literal
sense. When this condition is satisfied the values of $\Lambda $
in (\ref{LambdaM}) are real and the instability appears at any $
\vec q \neq 0$. [If one considers the inertial term in the
equation of motion (\ref{equaVec}), then even the uniform
deviations $U(t)$ from $U_0$ (dynamics with $\vec q = 0$) become
unstable when $dF_e/dV < 0$.]

Hence, the fulfillment of the condition $dF_e/dV < 0$ apparently
indicates the rise of the instability of the lattice translational
motion. But this condition is substantially more strict than
$F^{\prime}(V)<0$ (see Fig. 3), and as we will demonstrate in the
next Section never will be satisfied in our model (this condition,
though, could be met in other models with $dF/dV < 0$).

The second condition corresponds to the inequality:
\begin{equation}
\label{NC2} -F^\prime C_{yy} >(F/V)C_{xx}.
\end{equation}

The condition (\ref{NC2}) is one more sufficient condition of the
instability independent of (\ref{NC1}). Depending on the strength
of requirements (\ref{NC1},\ref{NC2}) one of them should be
fulfilled earlier than the other. We will demonstrate below that
the condition (\ref{NC2}) will be met in real superconductors
unlike the requirement (\ref{NC1}).

\section{Discussion}

In the previous section we have established a set of conditions of
the vortex lattice instability which comprises both the necessary
and sufficient requirements. Let us estimate the least strict
condition, the necessary condition $F^{\prime }(V)<0$. By
differentiating (\ref{sumdrag}) one obtains

\begin{equation}
\label{neccondbas}\frac \gamma {2\sqrt{V}}\frac{\left(
V/V_c-1\right) }{ \left( 1+V/V_c\right) ^2}>\beta.
\end{equation}

The left hand side of the inequality (\ref{neccondbas})  as a
function of the variable $V/V_c$  has the maximal value
0.04$\gamma/\sqrt V_c$ at $V/V_c \simeq 2$. Using $V_c=\eta
/\gamma $ one can rewrite (\ref{neccondbas}) as
\begin{equation}
\label{neccond}\beta \leq 0.04\cdot \gamma ^{3/2}/\sqrt{\eta }.
\end{equation}

The parameters $\eta$ and $\gamma$ can be taken from the limiting
formulae (\ref{limit1}-\ref{visccoef}) and expressed through
microscopic characteristics of the superconductor (see, e.g.,
Refs. \onlinecite{Blater}, \onlinecite{Tinkham}): the elastic
modulii  $ c_{11}=c_{44}=B^2/4\pi ,\ c_{66}=B\Phi _0/(8\pi \lambda
)^2$, where $B$ is the magnetic induction, $\Phi _0$ is the flux
quanta, $\lambda $ is the penetration depth. The vortex lattice
constant is determined by magnetic induction $a_v^2 =\left(
2/\sqrt{3}\right) \left( \Phi _0/B\right) $.  The Hall constant is
$ H=\Phi_0en/c$, where $e$ is the electron charge, $c$ is the
speed of light, $n$ is the density of superconducting electrons,
$n=mc^2/4\pi \lambda ^2e^2$ with the electron mass, $m$. The
parameter  $\alpha $ can be written as \cite{My+PRL} $\alpha
=a^3\left( H_{cm}^2/8\pi \right) \left( \Delta T_c/T_c\right)$,
where $H_{cm}$ is the superconducting thermodynamic critical
field, $\Delta T_c/T_c$ is a relative suppression of the critical
temperature nearly a defect, $a^3$ is the volume of the defect
($a$ is of the order of the interatomic distance). As a result, in
the case of the small dissipation, the Eq.(\ref{neccond}) takes
the following form:
\begin{equation}
\label{microcond} \frac{\beta}{H} \lesssim   10^{-2}\left(
C_{imp}a^3\right)\cdot \left( \frac{\Delta T_c}{T_c} \right) ^2
\frac{a^3\lambda ^3}{\xi ^6}
 \left(\frac{H_{c1}}{B}\right)^p,
 \end{equation}
where $H_{c1}=\Phi_0/4\pi\lambda^2$ is the quantity of the order
of the lower critical field and $p$ depends on $(c_{11}/c_{66})$
\begin{eqnarray}
p =
\begin{cases}
   19/8, &  c_{11}\gg c_{66}, \\
   11/4, & c_{11}\sim c_{66}.\\
  \end{cases}
  \end{eqnarray}

If one would take the numerical values of $\lambda $ and $\xi $,
as for typical high $T_{c}$ superconductor YBCO: \cite{Blater}
$\lambda =10^{-5}$cm, $\xi =10^{-7}$ cm, $a=5\cdot 10^{-8}$ cm,
$\Delta T_c/T_c\approx 1,C_{imp}a^3\approx 10^{-3}$ then
\begin{equation}
\label{microcond1}
     \left( H_{c1}/B\right )^p \geq \frac \beta H.
\end{equation}

This condition is met at magnetic fields close to the lower
magnetic field, $ H_{c1}$, and the microscopical parameters used
above for YBCO compound. The maximal values of $\beta =\Phi
_0^2/2\pi  \xi ^2c^2\rho _n$ , where $\rho _n$  is the normal
conductivity of the superconductor.  So the maximal value of the
ratio $\beta /H$ is $\beta /H=(1/\rho _n)(\lambda / \xi)^2(\hbar
/2\pi mc^2) $ and at $\rho _n=5\cdot 10^{-16}$ s (the typical
value for YBCO) it is equal approximately to $10^{-2}$. Note this
is just the upper limit of $\beta /H$ and for real superconductors
this value can be even smaller. Thus, for the case of low
viscosity the instability necessary condition may be fulfilled at
high enough values of magnetic fields.

In the case of high viscosity, $H\ll \beta $, the necessary
condition can be satisfied with difficulty as $\gamma (\beta)$
contains $\beta$ and the right hand side of the inequality
(\ref{neccondbas}) becomes unity (in conrast to the case $H \gg
\beta$ where the small parameter $\beta /H$ takes place).

Now we study the first sufficient condition $-F^{\prime
}(F/V)>H^2$. We carry out the calculation in the similar manner as
for (\ref{neccond}) and obtain (neglecting the bare dissipation as
$H\gg \beta $) that $0.1\cdot \gamma ^{3/2}/\sqrt{\eta }>H$.
Apparently this condition cannot be satisfied since though the
same inequality as (\ref{microcond1}) is valid for this condition,
but on the right hand side instead the small parameter $\beta /H$
unity is found. Then, we consider the second sufficient condition
$ -F^{\prime }/\left( F/V\right) >C_{xx}/C_{yy}$ which acquires
the following form:
\begin{equation}
\label{suffcondbus}\frac \gamma {2\sqrt{V}}\frac{\left( \left(
C_{yy}-2C_{xx}\right) V/V_c-\left( C_{yy}+2C_{xx}\right) \right)
}{\left( C_{yy}+C_{xx}\right) \left( 1+V/V_c\right) ^2}>\beta .
\end{equation}

It is seen that the inequality (\ref{suffcondbus}) has an obvious
resemblance with Eq.(\ref{neccondbas}). The difference is caused
by $\vec q$-dependent coefficients containing $C_{xx}$ and
$C_{yy}$ in Eq.(\ref{suffcondbus}).

For $2C_{xx} > C_{yy}$, which is realised, for instance, when
$c_{11}=c_{66}$, the left hand side of (\ref{suffcondbus}) is
negative. Therefore, for some values of $C_{xx},\ C_{yy}$ the
condition (\ref{suffcondbus}) cannot be fulfilled  and no
instability appears. However, by using the fact that $c_{66}
\simeq (c_{11}/4)(H_{c1}/B)\ll c_{11} $ and choosing an
appropriate $\vec q$ it can be satisfied. The simple analysis
shows that when $c_{11}\geq c_{66}$ the most favorable case is the
wave vectors along the $y$-axis, $\vec q \| \vec e_y$. Then, the
inequality (\ref{suffcondbus}) takes the following form:
\begin{equation}
\label{suffcondlim}\frac \gamma {2\sqrt{V}}\frac{\left( \left(
c_{11}-2c_{66}\right) V/V_c-\left( c_{11}+2c_{66}\right) \right)
}{\left( c_{11}+c_{66}\right) \left( 1+V/V_c\right) ^2}>\beta .
\end{equation}

The maximal value of left hand side of (\ref{suffcondlim}) is
attained at $3\left(c_{11}-2c_{66}\right)\left(V/V_c\right)=
\left( 3c_{11}+4c_{66}+2\bar c\right)$, where $\bar c = \sqrt
{3c_{11}^2+6c_{11}c_{66}+ c_{66}^2}$. It is equal to

\begin{equation}
\label{suffcondlim2}\frac {\gamma ^{3/2}}{\sqrt{\eta}}
\frac{(3\sqrt 3/4)\left(c_{11} -2c_{66}\right)^{5/2} \left(\bar c
-c_{66}\right)} {\left( c_{11}+c_{66}\right) \sqrt
{3c_{11}+4c_{66}+2\bar c}\left( 3c_{11} -c_{66}+\bar c\right) ^2}.
\end{equation}

Thus, the sufficient condition (\ref{suffcondlim}) is more strict
than necessary one given by (\ref{neccondbas}) owing to the
multipliers with $c_{11}$ and $c_{66}$. In particular, it can be
met only at $c_{11}>2c_{66}$. However, in the case of interest $
c_{11}\geq 10\  c_{66},$ the necessary and sufficient conditions
are close each other and the sufficient requirement
(\ref{suffcondlim}) can be satisfied as well. As the instability
is expected to arise when $H \gg \beta$ and $V\approx 2V_c$, one
can estimate $V_c$ for this case. In accordance with
Eq.(\ref{Vcrit}) and with the use of the above presented
microscopical parameters for YBCO $V_c$ can be presented as:
\begin{equation}
\label{estimate}
 V_c \approx 10^{-2}\cdot \left (B[\textrm{Oe}]\right)^{3/2},
\end{equation}
and at $B=10^3$ Oe, $V_c \approx 300$ cm/s. Thus, the instability
may appear under the rather strict condition (\ref{microcond1})
and at the velocities of the order of $2V_c \sim 10^3$ sm/s.
Hence, both the necessary and sufficient conditions can be
fulfilled at low viscosity, $\beta \ll H$, relatively small
magnetic fields and quite high velocities of the vortex lattice
motion.

At small velocities another effect, namely a nonlinear
\textit{I-V} characteristic of the superconductor emerges. This
effect does not impose limits for superconductor parameters. It
can be observed at $\beta > H$ and at strong magnetic fileds, when
$V_c \gg V_{trans}$ and no instability arises.

At $V_c \gg V_{trans}$, $F(V)$ dependence is monotonic  and
contains only part with $F \propto \sqrt V,\ V< V_{trans}, \ F
\propto V, \ V > V_{trans}$. In this case, the quadratic
\textit{I-V} characteristic is realised at $V < V_{trans}$. This
effect can be found at very small defect concentration and rather
dense lattice. For instance, if $ C_{imp}a^3=10^{-6},$
$V_{trans}\approx 10$ cm/s when $H \gg \beta$ and
$V_{trans}\approx 70$ cm/s for $H \ll \beta$ at $B=10^3$Oe.

\section{Conclusion}

We have demonstrated that due to the interaction of weak defects
with the Abrikosov vortex lattice the additional channel of
dissipation appears, which contributes to the total dissipation of
the moving lattice. As a result the additional friction force has
a non-linear dependence of the translational lattice velocity,
$F\propto \sqrt{V}$ for both weak bare dissipation, $H\gg \beta$,
and high viscosity $H\ll \beta $. This channel of dissipation
turns out to be an essential source of the system dissipation and
even prevails over the bare dissipation, $\beta V,$ below the
characteristic velocity $V_{trans}$.

The nonlinear character of the additional drag force results in a
nonlinear current-voltage characteristic of the superconductor.
Besides, the tangent of the Hall angle $\tan \alpha $ appears to
be the translational lattice velocity $V_0$ dependent up to the
characteristic velocity $V_{trans,H},$ above which it is
saturated.

Studying the stability of the vortex lattice against uniform and
nonuniform perturbations we have found several salient features.

The instability of the vortex lattice motion is found to occur in
the limit of small viscosity, $H\gg \beta $. We have revealed that
the negative differential mobility, $dV_0/dF_e$, is not
\textit{apriori} the lattice motion instability requirementl. Due
to the Hall component in the lattice motion, the value
$[dF(V)/dV]^{-1}$ does not coincide with the differential mobility
$dV/dF_{e}$. The estimates  suggest that condition $dV/dF_{e} < 0$
can be met with difficulty in contrast to $dV/dF < 0$. The
instability of the lattice motion regarding nonuniform
perturbations has been found when the value of (-$dF/dV$) is
greater than the combination of the elastic modulii $c_{11}$ and
$c_{66}$, while the more strict requirement $dV/dF_{e} < 0$ is
still not fulfilled. It is worth to note that the vortex lattice
instability concerned in this paper has different origin than
instabilities predicted in Ref. \onlinecite{LarOvch}.

\begin{acknowledgements}

The authors are indebted to V.~G.~Baryakhtar and A.~L.~Kasatkin
for stimulating discussions.  This work is supported by INTAS
Foundation grant No 97-31311.

\end{acknowledgements}

\eject

\begin{figure}
\includegraphics[width=3.5in]{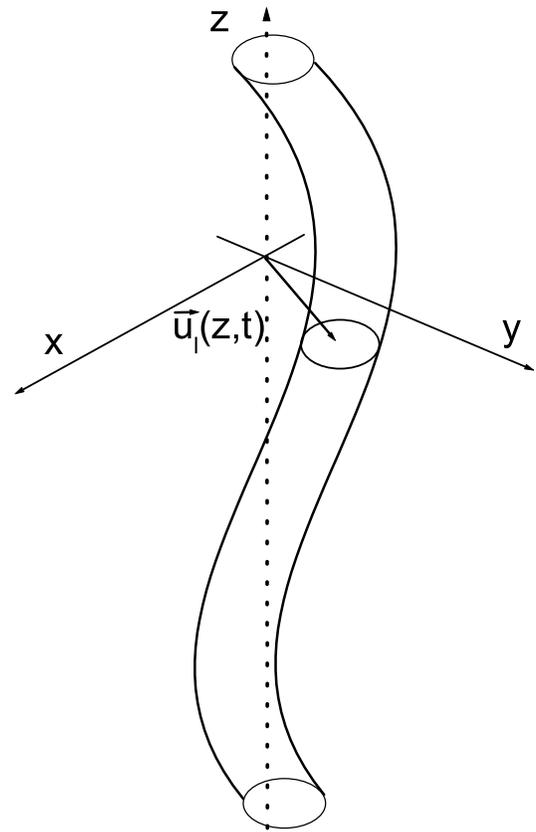}
\caption{The sketch of the vortex displacement $\vec u_l (z,t)$.}
\label{Fig1}
\end{figure}

\eject

\begin{figure}
\includegraphics[width=3.0in,angle=90.0]{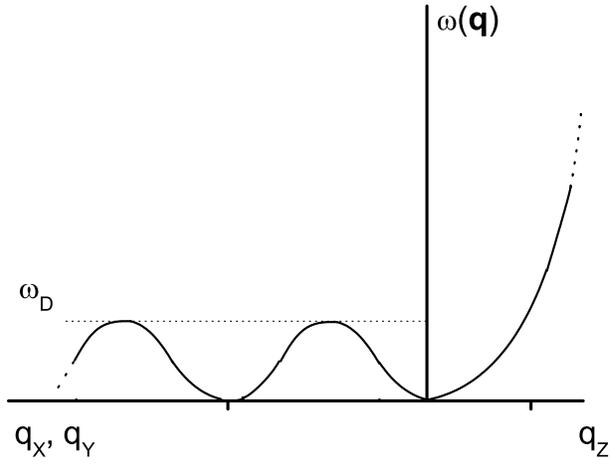}
\caption{The schematic representation of the dispersion law of the
vortex lattice. The dotted line indicates the Debye frequency,
$\omega_D$.} \label{Fig2}
\end{figure}

\eject

\begin{figure}
\includegraphics[width=3.5in]{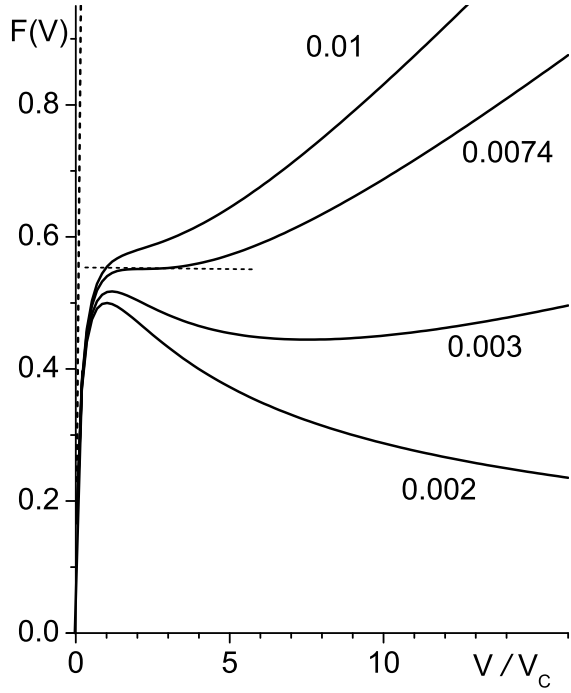}
\caption{The velocity dependencies of the drag force $F(V)$ at
$B=3H_{c1}$ for different values of the ratio $\beta/ H$ (its
value is shown near each curve). The dotted line parallel to the
$x$-axis marks the point on the curve with the critical ratio
$\beta/ H = 0.0074$ where the nonmonotonic character of $F(V)$
starts to develop. The dashed line describes the relation of the
external force, $F_e$ and the velocity $V$, at very small value of
the ratio $\beta/ H=0.002$. As was noted in the text the behaviour
of this curve is very far from nonmonotonic and can be hardly
distinguished from the $y$-axis.} \label{Fig3}
\end{figure}

\end{document}